\documentclass[a4paper,english,numberwithinsect,nosubfigcap]{eurocg24}

\usepackage{microtype} \usepackage{xspace}
\usepackage{todonotes}
\usepackage{cleveref}
\usepackage{thm-restate}
\theoremstyle{plain}

\newcommand{\syncplan}{{\normalfont\textsc{Syn}\-\textsc{chro}\-\textsc{nized} \textsc{Pla}\-\textsc{nari}\-\textsc{ty}}\xspace}
\newcommand{\cplan}{{\normalfont\textsc{Clus}\-\textsc{tered} \textsc{Pla}\-\textsc{nari}\-\textsc{ty}}\xspace}

\newcommand{\clplan}{{\normalfont{unrestricted} \textsc{Clus}\-\textsc{tered} \textsc{Level} \textsc{Pla}\-\textsc{nari}\-\textsc{ty}}\xspace}
\newcommand{\yclplan}{{\normalfont$y$-\textsc{mono}\-\textsc{tone} \textsc{Clus}\-\textsc{tered} \textsc{Level} \textsc{Pla}\-\textsc{nari}\-\textsc{ty}}\xspace}
\newcommand{\cclplan}{{\normalfont\textsc{Con}\-\textsc{vex} \textsc{Clus}\-\textsc{tered} \textsc{Level} \textsc{Pla}\-\textsc{nari}\-\textsc{ty}}\xspace}
\newcommand{\pmsat}{{\normalfont\textsc{Pla}\-\textsc{nar} \textsc{Mono}\-\textsc{tone} \textsc{3-SAT}}\xspace}
\newcommand{\shortclplan}{{\normalfont\textsc{uCLP}}\xspace}
\newcommand{\shortyclplan}{{\normalfont\textsc{$y$-CLP}}\xspace}
\newcommand{\shortcclplan}{{\normalfont\textsc{cCLP}}\xspace}

\graphicspath{{./graphics/}, {./graphics/pdf}}

\title{Clustered Planarity Variants for Level Graphs\thanks{Simon D. Fink was funded by
 the Deutsche Forschungsgemeinschaft (German Research Foundation, DFG) grant RU-1903/3-1
 and by the Vienna Science and Technology Fund (WWTF) [10.47379/ICT22029].}}
\titlerunning{Clustered Planarity Variants for Level Graphs}

\author[1,2]{Simon D. Fink}
\author[2]{Matthias Pfretzschner}
\author[2]{Ignaz Rutter}
\author[3]{Marie Diana Sieper}
\affil[1]{Algorithms and Complexity Group, Technische Universität Wien, Austria\\
  \texttt{sfink@ac.tuwien.ac.at}}
\affil[2]{Faculty of Computer Science and Mathematics, University of Passau, Germany\\
\texttt{\{finksim,pfretzschner,rutter\}@fim.uni-passau.de}}
\affil[3]{Institute of Computer Science, University of Würzburg, Germany\\
  \texttt{marie.sieper@uni-wuerzburg.de}}

\authorrunning{S.D. Fink, M. Pfretzscher, I. Rutter, M.D. Sieper}
\def\copyrightline{}

\begin{document}

\maketitle

\begin{abstract}
  We consider variants of the clustered planarity problem for
  level-planar drawings.  So far, only convex clusters have been studied in this setting.  We introduce two
  new variants that both insist on a level-planar drawing of the input
  graph but relax the requirements on the shape of the clusters.  In
  \clplan (\shortclplan) we only require that they are bounded by
  simple closed curves that enclose exactly the vertices of the
  cluster and cross each edge of the graph at most once.  The problem
  \yclplan (\shortyclplan) requires that additionally it must be
  possible to augment each cluster with edges that do not cross the
  cluster boundaries so that it becomes connected while the graph
  remains level-planar, thereby mimicking a classic characterization
  of clustered planarity in the level-planar setting.
  
  We give a polynomial-time algorithm for \shortclplan if the input graph
  is biconnected and has a single source.  By contrast, we show that
  \shortyclplan is hard under the same restrictions and it remains NP-hard
  even if the number of levels is bounded by a constant and there is
  only a single non-trivial cluster.
\end{abstract}

\section{Introduction}

A \emph{level graph} $(G,\gamma)$ is a graph~$G=(V,E)$ and a function~$\gamma \colon V \to \{1,2,\ldots,k\}$ with $k\in\mathbb{N}$ that assigns vertices to levels such that no two adjacent vertices are assigned to the same level.
A \emph{level planar drawing} of a level graph~$(G,\gamma)$ is a crossing-free drawing of~$G$ that maps each vertex~$v$ to a point on the line~$y = \gamma(v)$ and each edge to a $y$-monotone curve between its endpoints.
A level graph is \emph{level planar} if it has a level planar drawing.  Level planarity can be tested in linear time~\cite{jlm-lpt-98}.

Let~$G=(V,E)$ be a graph. A \emph{clustering}~$T$ of~$G$ is a rooted tree whose leaves are the vertices~$V$.
Each inner node~$\mu$ of~$T$ represents a \emph{cluster}, which encompasses all leaves~$V_\mu$ of the subtree rooted at~$\mu$. The pair~$(G,T)$ is called a \emph{clustered graph}.
A \emph{clustered planar drawing} of a clustered graph~$(G,T)$ is a planar drawing of~$G$ that also maps every cluster $\mu$ to a region~$R_\mu$ that is enclosed by a simple closed curve such that (i)~$R_\mu$~contains exactly the vertices $V_\mu$, (ii)~no two region boundaries intersect, and (iii)~no edge intersects the boundary of a cluster region more than once.  The combination of (i) and (iii) implies that an edge may intersect a cluster boundary if and only if precisely one of its endpoints lies inside the cluster.
A clustered graph is \emph{clustered planar} if it has a clustered planar drawing.
The problem of testing this property and finding such drawings is called \cplan.  In a recent breakthrough, Fulek and Tóth gave the first efficient algorithm for this problem~\cite{ft-aec-22}, which was soon after improved to a quadratic-time solution \cite{bfr-spw-21}.

In this paper, we seek to explore the combination of the two concepts of level planarity and clustered planarity.  Namely, our input is a \emph{clustered level graph (cl-graph)}, which is a tuple $(G,\gamma,T)$ such that $(G,\gamma)$ is a level graph and $(G,T)$ is a clustered graph.  We insist on a level-planar drawing of $G$.  However, it is not immediately clear  which conditions the cluster boundaries should fulfill.  Forster and Bachmaier~\cite{fb-clp-04} proposed the problem variant \cclplan (short \shortcclplan), which requires to draw the clusters as convex regions\footnote{they only considered this convex setting and used the name \textsc{CLP} instead of~\shortcclplan}.  They showed that \shortcclplan can be solved in linear time if the graph is proper (i.e., all edges connect vertices on adjacent levels) and the clusters are level-connected (i.e., each cluster contains an edge between any pair of adjacent levels it spans).  Angelini et al.~\cite{alb-tio-15} showed that testing \shortcclplan is NP-complete, but can be tested in quadratic time if the input graph is proper, thereby dropping the requirement of level-connectedness.

\begin{figure}
  \begin{subfigure}{.5\textwidth}
  \centering
  \includegraphics[page=1]{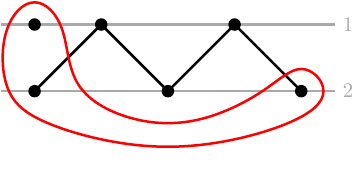}
  \caption{}
  \end{subfigure}\begin{subfigure}{.5\textwidth}
  \centering
  \includegraphics[page=2]{non-cclplan}
  \caption{}
  \end{subfigure}
  \caption{
    \textbf{(a)} A drawing that is level planar and clustered planar and thus cl-planar, but not convex cl-planar or $y$-cl-planar.
\textbf{(b)} A drawing that is $y$-cl-planar (with the augmentation edge in $E'$ shown dashed in red) and thus also cl-planar, but not convex cl-planar.
}
  \label{fig:non-cclplan}
\end{figure}

In this paper we consider two new variants that relax the conditions on the drawing of the cluster.
In~\clplan (short \shortclplan) we keep the conditions (i)--(iii) as stated above, i.e., the shapes of the clusters are not restricted by the levels.  Our second variant \yclplan (short \shortyclplan) is based on the characterization that a planar drawing~$\mathcal{G}$ of a graph~$G$ is clustered planar w.r.t.\ to a clustering~$T$ if and only if it is possible to insert a set of \emph{augmentation edges} into~$\mathcal{G}$ in a planar way such that each cluster becomes connected and no cycle formed by vertices of a cluster~$\mu$ encloses a vertex not in~$\mu$ in its interior~\cite{cw-ccc-06}. In analogy to this, we define a level-planar drawing to be \emph{$y$-cl-planar} if it satisfies these conditions but additionally the augmentation edges can be added as $y$-monotone curves.
\Cref{fig:non-cclplan} shows that \shortcclplan, \shortclplan, \shortyclplan are indeed different problems.  We are not aware of work that concerns \shortclplan or \shortyclplan.

We show that \shortclplan can be solved in polynomial time if the input graph is biconnected and has a single \emph{source}, i.e., a vertex that does not have a neighbor on a lower level.  On the other hand we show that \shortyclplan is NP-complete under the same conditions and also if the number of levels is 5 and there is only a single non-trivial cluster, i.e., that not contains all vertices.

\section{Single-Source Biconnected (unrestricted) Clustered Level Planarity}
\label{sec:algo}

We show that \shortclplan can be solved efficiently if~$G$ is a biconnected graph with a single source.  To this end, we combine the polynomial-time solution for~\cplan~\cite{bfr-spw-21} with a combinatorial description of all level-planar drawings of a biconnected single-source graph~\cite{br-ast-23}.

Note that whether a drawing of~$G$ is clustered planar depends only on its combinatorial embedding, rather than the precise drawing.  Thus, we call an embedding~$\mathcal E$ of~$G$ \emph{clustered planar} if the corresponding drawings are.  We call an embedding~$\mathcal E$ of $G$ \emph{level planar} if $G$ admits a level-planar drawing with embedding~$\mathcal E$.  To solve~$\shortclplan$ for an instance~$(G,\gamma,T)$, we need to find an embedding of~$G$ that is both cluster planar and level planar.

We first introduce yet another type of constraints called \emph{synchronized fixed-vertex constraints} (\emph{sfv-constraints} for short).  For a graph~$G=(V,E)$ an \emph{sfv-constraint} is a set~$Q$ of pairs $(v,\sigma_v)$, where~$v\in V$ and~$\sigma_v$ is a fixed cyclic order of the edges incident to~$v$, called its \emph{default rotation}.  An embedding~$\mathcal E$ of $G$ \emph{satisfies} the constraint~$Q$ if for each pair~$(v,\sigma_v) \in Q$ the rotation of~$v$ in~$\mathcal E$ is its default rotation or if for each pair~$(v,\sigma_v) \in Q$ the rotation of $v$ in~$\mathcal E$ is the reverse of its default rotation.  
Given a set~$\mathcal Q$ of sfv-constraints, we say that an embedding of~$G$ \emph{satisfies}~$\mathcal Q$ if it satisfies each~$Q \in \mathcal Q$. 

We use sfv-constraints to bridge from level-planarity to usual planarity.  To this end, we introduce a slightly generalized version of \cplan, where we seek a clustered-planar embedding that additionally satisfies a given set~$\mathcal Q$ of sfv-constraints.  This problem is called {\sc Sync CP}.  The point is that the algorithm of Bläsius et al.\ \cite{bfr-spw-21} reduces \cplan to the intermediate problem \syncplan, which includes the option to directly express sfv-constraints.

\begin{restatable}{fixlemma}{cplanfixedlem}
  \label{lem:cplan-fixed-constraints}
  {\sc Sync CP} can be solved in~$O(n^3)$ time.
\end{restatable}

\begin{figure}
  \centering
  \begin{subfigure}{.5\textwidth}
    \includegraphics[page=1]{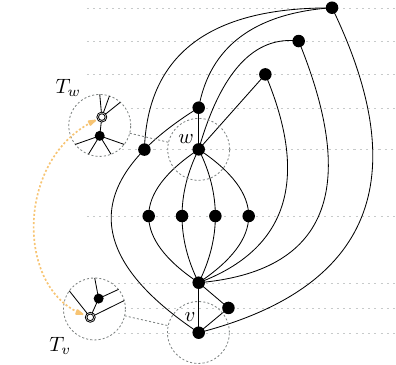}
    \caption{}
    \label{fig:LPTreeA}
  \end{subfigure}
  \begin{subfigure}{.45\textwidth}
    \includegraphics[page=2]{LPTree}
    \caption{}
    \label{fig:LPTreeB}
  \end{subfigure}
  
  \begin{subfigure}{0.5\textwidth}
    \includegraphics[page=3]{LPTree}
    \caption{}
    \label{fig:LPTreeC}
  \end{subfigure}
  \caption{\textbf{(a)} A level graph~$G$ with two level PQ-trees $T_w$ and $T_v$ derived from \textbf{(b)} its LP-tree. P-nodes are represented by black disks, Q-nodes as white double disks. \textbf{(c)} The graph after replacing $w,v$ by $T_w,T_v$;  the orange arrow indicates the sfv-constraint due to $\rho$.}
  \label{fig:LPTree}
\end{figure}

We now turn to the second ingredient.  Let~$(G,\gamma)$ be a biconnected single-source level-planar graph and let~$\Gamma$ be an arbitrary level-planar embedding of $G$.  Brückner and Rutter~\cite{br-ast-23} showed that there exists a data structure, called LP-tree, very similar to the famous SPQR-tree, that represents precisely the level-planar embeddings of $G$; see \cref{fig:LPTree} for an example.  Like the SPQR-tree, the embeddings decisions for the LP-tree are made by (i) arbitrarily reordering parallel subgraphs between a pair of vertices and (ii) flipping the embeddings of some disjoint and otherwise rigid structures.  Hence, the possible orderings of the edges around each vertex~$v$ of $G$ in any level-planar embedding can be described by a PQ-tree~\cite{bl-tft-76, fpr-eco-21}~$T_v$, called \emph{level PQ-tree} that is straightforwardly derived from the LP-tree; it contains one P-node~$u_{v, \mu}$ for each parallel structure~$\mu$ in which~$v$ occurs and one Q-node~$u_{v,\rho}$ for each rigid structure~$\rho$, in which~$v$ occurs; see \cref{fig:LPTreeA}.  The level-planar embedding~$\Gamma$ is used as reference to determine a \emph{default rotation}~$\sigma_{v,\rho}$ for each Q-node~$u_{v,\rho}$.

If an embedding~$\mathcal E$ of $G$ is level-planar, the rotation of each vertex~$v$ is necessarily represented by its level PQ-tree~$T_v$.  It further holds that all Q-nodes~$u_{v,\rho}$ with~$v \in V$ that stem from the same rigid structure either all have their default orientation or all its reversal.   An embedding where the last condition holds for each rigid structure is called~\emph{$\rho$-consistent}.

\begin{restatable}{fixlemma}{levelcharacterizationlem}
  \label{lem:level-characterization}
  An embedding $\mathcal E$ is level-planar if and only if the
  rotation of each vertex~$v$ is represented by its level
  PQ-tree~$T_v$ and moreover~$\mathcal E$ is~$\rho$-consistent.
\end{restatable}

For a biconnected single-source level-planar graph~$G$ with level-planar embedding~$\Gamma$, we derive a new graph~$G^+$ and a set~$\mathcal Q$ of synchronized fixed-vertex constraints.  We replace each vertex~$v$ of~$G$ by a tree isomorphic to its level PQ-tree~$T_v$; see \Cref{fig:LPTreeC}.  In order to enforce $\rho$-consistency, we additionally create for each rigid structure~$\rho$ of the LP-tree an sfv-constraint~$Q_\rho = \{(u_{v,\rho}, \sigma_{v,\rho}) \mid $ $v$ occurs in the rigid structure~$\rho \}$.  Let~$\mathcal Q$ denote the set of these constraints for all rigid structures.  Clearly, we can obtain a planar embedding of~$G$ by taking a planar embedding of~$G^+$ that satisfies~$\mathcal Q$ and contracting each tree~$T_v$ back into a single vertex.  The embeddings we can obtain in this way are precisely those where the rotation of each vertex~$v$ is represented by its level PQ-tree~$T_v$ and that are~$\rho$-consistent.

Finally, it is time to connect clusters and level planarity.  To this end, consider a clustering~$T$ on~$G$.  We naturally obtain a corresponding clustering~$T^+$ of~$G^+$ by placing each vertex of~$G^+$ into the cluster of the vertex of~$G$ it replaces.

\begin{restatable}{fixlemma}{cplancorrectnesslem}
  \label{lem:clplan-correctness}
  $(G,\gamma,T)$ admits a planar embedding that is level-planar and clustered-planar if and only if~$(G^+,T^+)$ admits a clustered-planar embedding that satisfies~$\mathcal Q$.
\end{restatable}

Altogether, this reduces the problem \shortclplan of biconnected single-source graphs to {\sc Sync CP}, which can be solved efficiently by \Cref{lem:cplan-fixed-constraints}.

\begin{restatable}{theorem}{ssbicoclplanthm}\label{sec:ssbicoclplanthm}
  \shortclplan can be solved in $O(n^3)$ time for biconnected single-source level graphs.
\end{restatable}

\section{Hardness of $y$-monotone Clustered Level Planarity}
\label{sec:complexity}

It is easy to see that \shortyclplan lies in NP by guessing and verifying the augmentation edges and an embedding. We show that it is NP-hard even for inputs with very restricted properties.

\begin{figure}
  \centering
  \begin{subfigure}[t]{0.45\textwidth}
    \includegraphics[page=1]{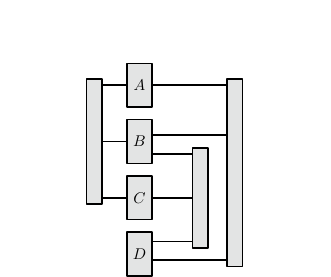}
    \caption{}
    \label{fig:incidenceGraphA}
  \end{subfigure}
  \hfill
  \begin{subfigure}[t]{0.45\textwidth}
    \includegraphics[page=2]{incidenceGraph}
    \caption{}
    \label{fig:incidenceGraphB}
  \end{subfigure}
  \begin{subfigure}[t]{\textwidth}
    \includegraphics[]{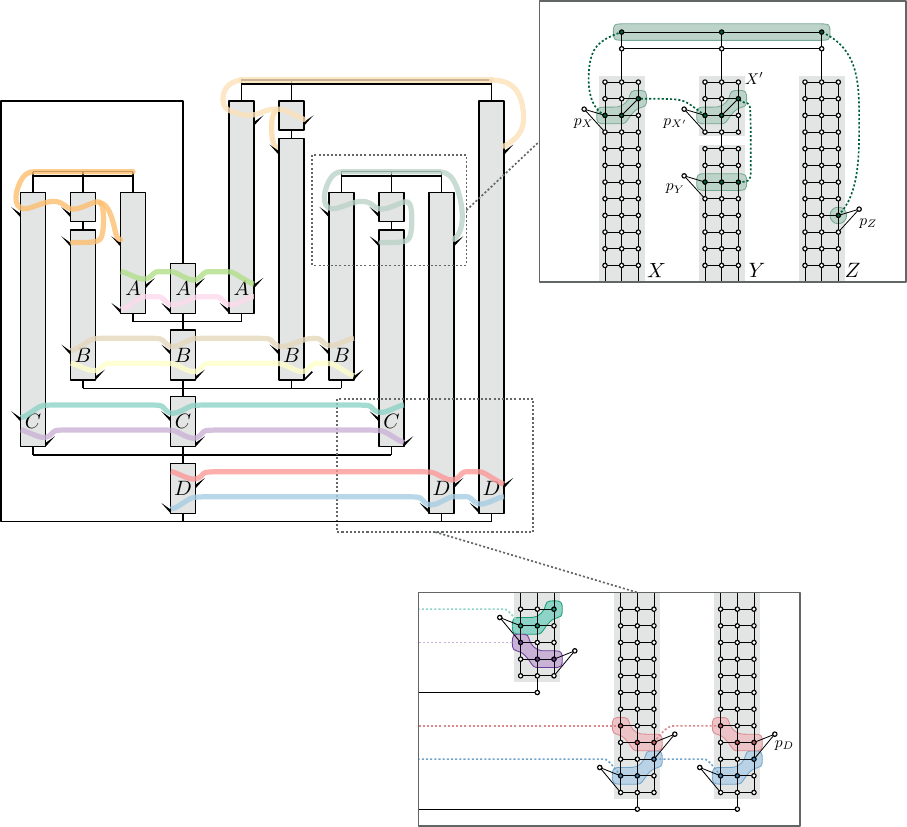}
    \caption{}
    \label{fig:PM3SatOverview}
  \end{subfigure}
  \caption{\textbf{(a)} An instance of \pmsat. \textbf{(b)} The modified incidence graph. 
    \textbf{(c)}~The structure of the corresponding \shortyclplan instance. 
    Highlighted are a clause gadget (top right) and the gadget for propagating variable assignments (bottom right).}
  
  \label{fig:PM3SatReduction}
\end{figure}

\begin{figure}
  \centering
  \begin{subfigure}[t]{0.45\textwidth}
    \includegraphics[scale=1]{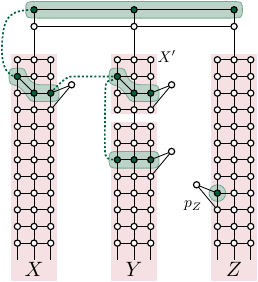}
    \caption{}
    \label{fig:PM3SatClauseGadgetInvalidA}
  \end{subfigure}
  \hfill
  \begin{subfigure}[t]{0.45\textwidth}
    \includegraphics[scale=1,page=2]{PM3SatClauseGadgetInvalid}
    \caption{}
    \label{fig:PM3SatClauseGadgetInvalidB}
  \end{subfigure}
  \caption{Neither flip of~$X'$ admits a valid embedding of the clause gadget if all literals are false.}
  \label{fig:PM3SatClauseGadgetInvalid}
\end{figure}

\begin{figure}
  \centering
  \includegraphics[scale=1]{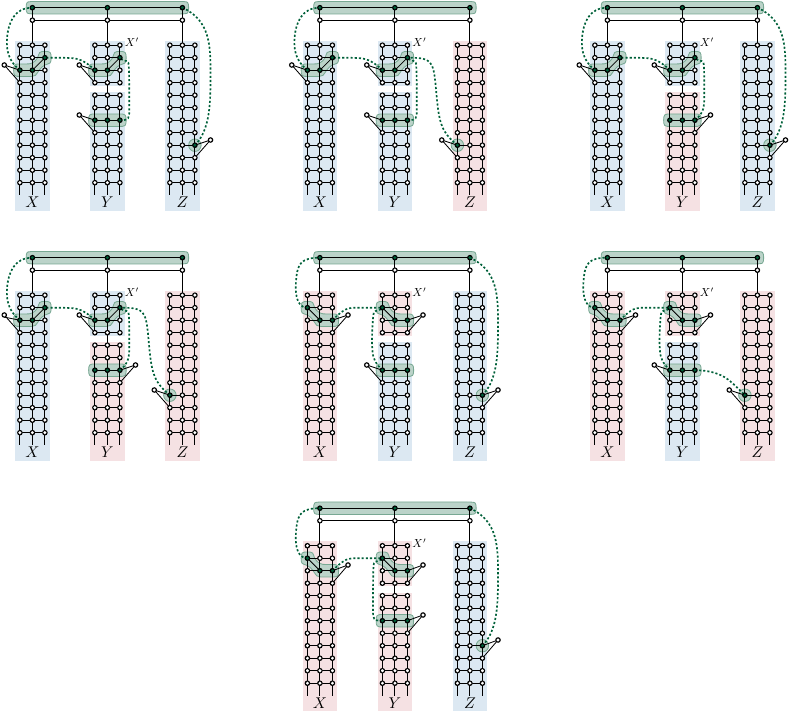}
  \caption{The seven variable assignments for which the clause gadget admits a valid embedding.}
  \label{fig:PM3SatClauseGadgetValid}
\end{figure}

\begin{restatable}{theorem}{pmsatthm}
  \yclplan is NP-complete, even if the input is a biconnected graph with just one source. \end{restatable}

\begin{proof}[Proof Sketch]
  We reduce from the NP-complete problem \pmsat~\cite{bk-obs-12}, which asks for the satisfiablity of {\sc 3-SAT} formulas whose incidence graph has a drawing where all 
variables lie on a vertical line~$\ell$, each clause contains only positive or only negative literals, and the positive and negative clauses lie on opposing sides of~$\ell$; see \Cref{fig:incidenceGraphA}. 
  
  Given such a drawing for a formula~$\phi$, we first reorient the clauses horizontally above the variables as illustrated in \Cref{fig:incidenceGraphB}.
  From this drawing, we construct an equivalent instance of \shortyclplan as illustrated in \Cref{fig:PM3SatOverview}.
  The variables and literals are represented by triconnected pillars (a $(3 \times k)$-grid for suitable $k$) that extend vertically towards the clause gadgets.
  The horizontal flip of a pillar represents its truth value.
   To synchronize pillars of the same variable, we use wedges that enclose a vertex of a disconnected cluster and, due to the required y-monotonicity for cluster connections, prohibit corresponding wedges of adjacent pillars to face each other, as such a connection would have to bypass both wedges.
  Using two clusters per variable, we can thus ensure consistent flips for all variables; see \Cref{fig:PM3SatOverview}.
  
  Using a similar approach with wedges, we can construct a clause gadget that allows all assignments for its literals except when all three literals are false; see \Cref{fig:PM3SatOverview} for the structure of the gadget.  As shown in \Cref{fig:PM3SatClauseGadgetInvalid}, there exists one configuration for the literal pillars of a clause where no valid embedding is possible, as the cluster of the gadget cannot connect with only y-monotone curves. \Cref{fig:PM3SatClauseGadgetValid} shows valid embeddings for all other configurations.
This way, we can construct in polynomial time an equivalent instance of \shortyclplan where the graph is biconnected and only has a single source.
\end{proof}

Our second reduction is from \textsc{3-Partition}, whose input is a multiset $A = \{a_1, \dots, a_{3m}\}$ positive integers and a bound $B \in \mathbb{N}^+$ with $B/4 < a_i < B/2$ and $\sum_{a \in A} a = m \cdot B$. The question is whether~$A$ can be partitioned into~$m$ sets $A_1, A_2, \dots, A_m$, such that for every~$j \in \{1, \dots, m\}$ it is $\sum_{a \in A_j} a = B$.
\textsc{3-Partition} is strongly \textsc{NP}-complete, i.e., it remains \textsc{NP}-complete even if 
$B$ is polynomial in $m$~\cite{doi:10.1137/0204035}.

\begin{restatable}{theorem}{threepartthm}
\yclplan is NP-complete, even if the input contains only one non-trivial cluster, the number of levels is at most 5, and all vertices have a fixed rotation.
\end{restatable}

\begin{proof}[Proof Sketch]
\begin{figure}
   \begin{subfigure}{.55\textwidth}
    \includegraphics[]{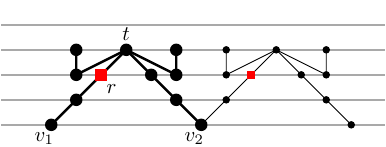}
    \label{fig:receiver}
    \caption{}
   \end{subfigure}
   \begin{subfigure}{.35\textwidth}
    \includegraphics[]{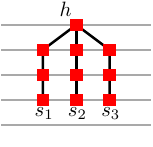}
    \label{fig:plug}
    \caption{}
   \end{subfigure}
   \begin{subfigure}{\textwidth}
    \includegraphics[]{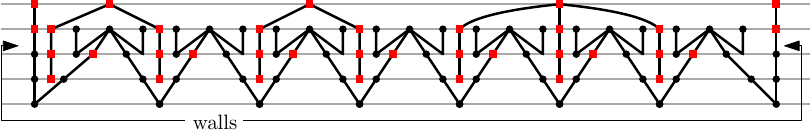}
    \label{fig:bucket}
    \caption{}
   \end{subfigure}
    \caption{\textbf{(a)} A receiver (bold) with a marked vertex~$r$ (box), and a second receiver (non-bold) chained to the first one.
    \textbf{(b)} A 3-plug.
    \textbf{(c)} A bucket of size 7, filled with two 2-plugs and a 3-plug. Every marked bucket vertex can be connected to a pin with a $y$-monotone curve.}
    \label{fig:3part-gadgets}
\end{figure}

Let~$(m, A, B)$ be an instance of \textsc{3-Partition}. We construct an instance of \shortclplan with a single non-trivial cluster~$\mu$. The main idea is to build $m$ \emph{buckets} (the structure in \Cref{fig:3part-gadgets}c) by chaining $B$ \emph{receivers} (the structure in \Cref{fig:3part-gadgets}a, each of which contains a \emph{connector vertex}~$r$ that belongs to~$\mu$), and closing the sides of each bucket with two paths (also called \emph{walls}).
Note that each of the $B$ connector vertices of a bucket must be connected to the rest of~$\mu$ due to the paths of length~2 attached to the vertices marked as~$t$ in the receiver; see~\Cref{fig:3part-gadgets}a.
For every $a \in A$, we generate an \emph{$a$-plug}; the structure in \Cref{fig:3part-gadgets}b). The leaves of a plug are called \emph{pins} and these are the only vertices that can link the connector vertices to the remainder of cluster~$\mu$.
Given a plug and a bucket, either all of the vertices of the plug are drawn between the two bucket walls, or none of them. Thus, we can model a solution $A_1, \dots, A_m$ with the~$m$ buckets, and a drawing assigns~$a\in A$ to~$A_i$ if and only if the corresponding $a$-plug is in the $i$-th bucket.
Since every pin can join at most one connector vertex, there are at least $B$ pins inside a bucket in a valid drawing. 
Since there are $m$ buckets and a total of $m\cdot B$ pins, the instance is valid if and only if we can distribute the plugs in such a way that 
there are precisely~$B$ pins per bucket, which corresponds directly to a solution of~$(m, A, B)$.
\end{proof}

\section{Conclusion}

We have introduced the problems \clplan and \yclplan, gave an polynomial-time algorithm for \clplan if restricted to biconnected single-source graphs, and showed that \yclplan is NP-complete under very restricted conditions.

We conclude by providing some open questions.
On the one hand, these are inspired by the restrictions imposed by our algorithm.
The LP-trees we use in \Cref{sec:ssbicoclplanthm} only exist for biconnected single-source instances and it is unlikely that this concept can be extended to multiple sources \cite[Section 5]{br-ast-23}.
Is it possible to extend our algorithm to non-biconnected graphs?
More generally, what is the complexity of \clplan?
On the other hand, the NP-hardness results on \yclplan and (non-proper) \cclplan \cite{alb-tio-15} raise the question whether these problems are FPT with respect to natural parameters.

\bibliography{references}

\newpage

\appendix

\section{Omitted Proofs from Section~\ref{sec:algo}}

\cplanfixedlem*

\begin{proof}
  We show the stronger statement that an instance $(G,T,Q)$ of {\sc Sync CP} can be solved in time $O(n+d\cdot\Delta + \sum_{Q \in \mathcal Q}|Q|)$ time, where $d$ is the total number of crossings between an edge and a cluster boundary and $\Delta$ is the maximum number of such crossings on a single cluster boundary.
  The planarity of the input graph guarantees that $m \in O(n)$ and thus $d \in O(n^2)$ and $\Delta \in O(n)$. 
  Because further $\sum_{Q \in \mathcal Q} |Q|$ is in~$O(n)$, the statement of the Lemma follows.
  
  We will first summarize the linear reduction of Bläsius et al.~\cite{bfr-spw-21} from \cplan to \syncplan to later show that it can easily also accomodate the synchronized fixed-vertex constraints.
  We obtain the \emph{skeleton} graph $G_\nu$ of the root cluster $\nu$ from the input graph $G$ by contracting all vertices in any child cluster $\mu$ of $\nu$ into a single vertex $v_\mu$.
  We can similarly obtain the skeleton $G_\mu$ of any non-root cluster $\mu$ by not only individually contracting all its respective child clusters, but also contracting all vertices $V\setminus V_\mu$ (that is, those not in $\mu$) into a vertex $v_{\overline\mu}$.
  Note that, for each non-root cluster $\mu$, there is one vertex $v_{\overline\mu}$ in its skeleton $G_\mu$ into which we contracted $V\setminus V_\mu$ and one vertex $v_\mu$ in another skeleton into which we contracted $V_\mu$.
  A datastructure labeling each cluster in $T$ with its skeleton was introduced under the name CD-tree by Bläsius et al.~\cite{br-anp-16}.

  The \syncplan instance corresponding to $(G,T)$ is based on the graph $G'$ comprising the union of all skeletons.
  \syncplan allows us to specify a set of \emph{pipes} $P$ which is a matching on a subset of the vertices of $G'$.
  Any valid embedding will have the rotation of one pipe endpoint vertex being the reverse of the rotation of the other pipe endpoint it is matched with under a bijection given with the pipe.
  For each non-root cluster $\mu$, we add a pipe between $v_{\mu}$ and $v_{\overline\mu}$ to $P$.
  The bijection naturally arises from the fact that the edges incident to $v_{\mu}$ and $v_{\overline\mu}$ stem from the same edges in $G$, which are those between $V_\mu$ and $V\setminus V_\mu$.

  \syncplan additionally allows for subsets of the vertices to be assigned a default rotation.
  Analogously to synchronized fixed-vertex constraints, in a valid \syncplan embedding, all vertices in the same such set are required to either all have their default rotation or all have the reverse of their default rotation.
  As $G'$ includes all vertices from $V$, we can simply formulate our synchronized fixed-vertex constraints $Q$ as what is called \emph{Q-constraints} in \syncplan.  The main difference is that in \syncplan, each vertex may be part of only one $Q$-constraint. This can be established in~$O(\sum_{Q \in \mathcal Q}|Q|)$ time by merging sets that constrain a common vertex. (If the same vertex is constrained with default rotations that are neither the same nor the reverse of each other, we can immediately conclude that this is a no-instance.)
  
  Alltogether, we obtain an instance $(G',P,Q)$ of \syncplan,\footnote{Note that in the original definition of \syncplan~\cite{bfr-spw-21}, the vertex sets and default rotations of $Q$ are given as separate variables. To keep notation consistent, we will not make this distinction here.} for which we will now show equivalence with the input \cplan {\sc~with Synchronized Fixed-Vertex Constraints} instance $(G,T,Q)$.

  Let $\mathcal E$ be a valid embedding of $(G,T,Q)$.
  We show that there is a valid embedding $\mathcal E'$ of the instance $(G',P,Q)$.
  As $\mathcal E$ is a c-planar embedding of $(G,T)$, there is a c-planar drawing $\mathcal{G}$ of $(G,T)$ that uses the rotations of $\mathcal E$.
  In this drawing, we can contract the region $R_\mu$ representing cluster $\mu$ (and thereby also its vertices $V_\mu$) into a single point $v_\mu$ while maintaining planarity as well as the rotations of all remaining vertices.
  Similarly, we can also contract the outside of any region $R_\mu$ (and thereby the vertices $V\setminus V_\mu$) into a single point $v_{\overline\mu}$, again maintaining planarity and vertex rotations.
  In this way, we can obtain a planar drawing and therefore also a planar embedding for each skeleton $G_\mu$.
  Combining these embeddings yields and embedding $\mathcal E'$ of $G'$.
Note that as $\mathcal E$ is c-planar and thus the cyclic orders in which edges cross a cluster region boundary are the same on both sides of the boundary, the rotations of $v_{\mu}$ and $v_{\overline\mu}$ are the reverse of each other for every $\mu$.
  Thus, the constraints arising from the pipes $P$ are satisfied by $\mathcal E'$.
  Furthermore, note that we did not change the rotation of any vertex of $V$ in $\mathcal E'$.
  As the set of vertices constrained by the synchronized fixed-vertex constraints $Q$ is a subset of $V$, the Q-constraints $Q$ are also satisfied in the \syncplan embedding $\mathcal E'$.

  Conversely, let $\mathcal E'$ be a valid embedding of the \syncplan instance $(G',Q,P)$.
  We show there is a valid embedding $\mathcal E$ of the corresponding instance $(G,T,Q)$ of \cplan {\sc with Synchronized Fixed-Vertex Constraints}.
  Let $\mathcal E'_\nu$ be the planar sub-embedding of $\mathcal E'$ for the skeleton $G_\nu$ of the root cluster $\nu$.
  Let $\mathcal{G}'_\nu$ be a drawing of $\mathcal E'_\nu$ and similarly, let $\mathcal{G}'_\mu$ be a drawing of a child cluster $\mu$ of $\nu$. Consider a disk $R_\mu$ around $v_\mu$ in $\mathcal{G}'_\nu$ that only contains (parts of) the edges incident to $v_\mu$.
  Similarly, consider a disk $R_{\overline\mu}$ around $v_{\overline\mu}$ in $\mathcal{G}'_\mu$.
  The $v_\mu$-$v_{\overline\mu}$ pipe defines a bijection $\varphi$ between these incident edges, and the cyclic order under which these edges intersect $R_\mu$ and $R_{\overline\mu}$ are the reverse of each other.
  We can therefore replace the inside of $R_\mu$ with the outside of $R_{\overline\mu}$ and connect edges intersecting the disks according to $\varphi$ while mainting planarity.
Effectively, this inserts $\mathcal{G}'_\mu$ instead of $v_\mu$ in $\mathcal{G}'_\nu$ and thereby undoes the contraction of $V_\mu$.
  Recursively applying this procedure for all non-root clusters undoes all contractions and thus yields a drawing $\mathcal{G}$ of $G$.
  The drawing is planar as the individual drawings are planar and the orders of the identified edges are the reverse of each other.
  The drawing is also c-planar when using $R_\mu$ as region for cluster $\mu$.
  Thus, we obtain a c-planar embedding $\mathcal E$ of $G$.
  It only remains to show the synchronized fixed-vertex contraints $Q$ are satisfied, which again directly follows from the rotations of any vertex in $V$ remaining unchanged by the performed insertions.

  This concludes the proof that we obtain an equivalent instance of \syncplan for any instance of \cplan {\sc with Synchronized Fixed-Vertex Constraints}.
  The size of the resulting \syncplan instance is bounded by $O(n+d)$~\cite{bfr-spw-21} and it can be solved in $O(n+d\cdot\Delta)$ time~\cite{fr-mtc-23}, where $d$ is the total number of crossings between an edge and a cluster boundary and $\Delta$ is the maximum number of such crossings on a single cluster boundary.
\end{proof}

\levelcharacterizationlem*

\begin{proof}
  The necessity is clear, the sufficiency follows from the fact that the conditions ensure that the planar embedding~$\mathcal E$ determines either the same embedding of the rigid structure~$\rho$ as the level-planar embedding~$\Gamma$ or its reverse.  It therefore follows that an embedding that satisfies these additional conditions is represented by the LP-tree and is therefore level-planar~\cite{br-ast-23}.
\end{proof}

\cplancorrectnesslem*

\begin{proof}
  Let~$\mathcal E$ be an embedding that is both level-planar and clustered planar.  By \Cref{lem:level-characterization}, since~$\mathcal E$ is level-planar, the rotation around each vertex~$v$ is represented by the corresponding LP-tree~$T_v$.  We can therefore perform the replacement of the vertices in the construction of~$G^+$ also in the embedding~$\mathcal E$, which thereby defines a unique planar embedding~$\mathcal E^+$ of~$G^+$.  Since this expansion can be done in a drawing that includes the cluster boundaries without introducing edge--cluster crossings that are not allowed, the resulting embedding~$\mathcal E^+$ is clustered planar.  Again by \Cref{lem:level-characterization} the fact that~$\mathcal E$ is level-planar further implies that~$\mathcal E$ is $\rho$-consistent and hence that~$\mathcal E^+$ satisfies~$\mathcal Q$.

  Conversely, assume that~$\mathcal E^+$ is a clustered-planar embedding of~$(G^+,T^+)$ that satisfies~$\mathcal Q$.  Let~$\mathcal E$ be the embedding obtained from~$G$ by contracting each of the trees~$T_v$ back into a single vertex~$v$.  Since we only contract edges whose endpoints lie in exactly the same clusters, it follows that~$\mathcal E$ is clustered planar.  Moreover, since the rotation around each vertex~$v$ of~$G$ in~$\mathcal E$ is obtained by contracting an embedding of~$T_v$ in~$\mathcal E^+$, it follows that the rotation of~$v$ in~$\mathcal E$ is represented by its level PQ-tree~$T_v$.  Finally, the fact that~$\mathcal E^+$ is level-planar further implies that~$\mathcal E$ is $\rho$-consistent.  By \Cref{lem:level-characterization} it follows that~$\mathcal E$ is level planar.
\end{proof}

\ssbicoclplanthm*

\begin{proof}
  Let~$(G,T,\gamma)$ be a clustered biconnected single-source level graph  and let~$(G^+,T^+), \mathcal Q$ denote the clustered graph and the synchronized fixed-vertex constraints constructed above.
  By \Cref{lem:clplan-correctness}, $(G,T,\gamma)$ is clustered level-planar if and only if $(G^+,T^+)$ admits a clustered-planar embedding that satisfies~$\mathcal Q$.  The latter is a simple instance of \cplan {\sc~with Synchronized Fixed-Vertex Constraints}, which can be solved efficiently by~\Cref{lem:cplan-fixed-constraints}.

  For the running time, observe that the LP-tree for~$(G,\gamma)$ can be computed in linear time~\cite{br-ast-23}.  The total size of all level PQ-trees $T_v$ of~$G$ is~$O(n)$ and they can be computed in the same running time.  We can thus compute~$(G^+,T^+)$ and also the constraints~$\mathcal Q$ in linear time.  Then the running time follows from \Cref{lem:cplan-fixed-constraints}.
\end{proof}

\section{Omitted Proofs from Section~\ref{sec:complexity}}

\pmsatthm*

\begin{proof}
  We reduce from the NP-complete problem \pmsat~\cite{bk-obs-12}, which is defined as follows.
  
  A \textsc{3-SAT} formula is \emph{monotone} if every clause is  either \emph{positive} or \emph{negative}, i.e., it contains only positive or only negative literals, respectively.
  The \emph{incidence graph} of a \textsc{3-SAT} formula is a bipartite graph between its clauses and variables, where the edges correspond to the occurrences of variables in the clauses.
  In a \emph{monotone rectilinear drawing} of the incidence graph, all variables and clauses are represented by rectangles, the variables lie on a vertical line and the positive and negative clauses lie on opposing sides of this line.
  The edges are represented by horizontal straight-line segments.
  Given as input a monotone \textsc{3-SAT} formula $\phi$ together with a planar monotone rectilinear drawing $\mathcal E$ of its incidence graph, \pmsat asks whether $\phi$ is satisfiable.
  
  Given an instance $(\phi, \mathcal E)$ of \pmsat, we construct an equivalent instance of \shortyclplan as follows.
  \Cref{fig:PM3SatReduction} shows the overall structure of the resulting instance, as well as the gadgets for clauses and for propagating variable assignments.
  In our construction, we implicitly allow edges between vertices on the same level, as these can be easily eliminated by adding intermediate levels and distributing the adjacent vertices among them.
  
  We begin by generating a new rectilinear drawing $\mathcal{E'}$ out of $\mathcal E$ by reorienting the clause rectangles horizontally, placing the clauses above the variables; see \Cref{fig:incidenceGraphB}.
  This introduces a bend in every edge.
  We now construct a clustered level graph out of $\mathcal{E'}$.
  
  The variables of $\phi$ are represented by triconnected pillars (a $(3\times k)$-grid for suitable~$k$) that are placed on a vertical line, where the horizontal flip of the pillar represents the corresponding truth value.
  Each variable is associated with a dedicated horizontal area containing pillars that represent the corresponding literals of the variable.
  The pillars corresponding to positive literals lie to the right of the variable, while the negative literals lie to the left.
  These pillars extend to the top where they join in their respective clause gadgets.
  
  To synchronize the rotation of all pillars corresponding to the same variable, we use the following technique.
  For each pillar $D$ that requires synchronization, we add a degree-2 vertex $p_D$, whose level lies above that of its neighbors, to the side of $D$, forming a wedge towards the top; see \Cref{fig:PM3SatOverview}.
  The level of $p_D$ is chosen high enough such that the wedge must always lie on the outer face of $D$.
  Moreover, we add a cluster $H$ that spans both sides of the pillar (red cluster in \Cref{fig:PM3SatOverview}), such that the end of $H$ on the side of $p_D$ lies within the wedge, and the end on the other side lies above $p_D$.
  The wedge, together with the required y-monotonicity, ensures that the vertex of $H$ it encloses can only connect to vertices that lie above $p_D$.
  Therefore, if we have two adjacent pillars that contain this structure, the sides containing the wedges cannot face each other, as otherwise, the vertices of cluster $H$ cannot connect.
  Repeating this construction using wedges on the other side of the pillars and a different cluster (blue cluster in \Cref{fig:PM3SatOverview}), we can thus synchronize the rotation of pillars.
  We use this technique to synchronize all literals corresponding to the same variable using two clusters.
  
  For every clause $c$ of $\phi$, the corresponding gadget is structured as follows.
  Let $x$, $y$ and $z$ denote the three variables contained in $c$, in the order induced by their appearance in $c$ in $\mathcal E$.
  We let $X$, $Y$, and $Z$ denote the corresponding pillars.
  Pillar $Y$ is further subdivided such that the topmost portion $X'$ can rotate independently to mimic and propagate the configuration of~$X$.
  Note that the clause gadget should yield a valid embedding of \shortyclplan for all variable assignments except one.
  We ensure this using vertices of a disconnected cluster $H_c$ (green vertices in \Cref{fig:PM3SatOverview}) that seek to connect to one another.
  At the top of the clause gadget, there are connected vertices of $H_c$, allowing vertices of $H_c$ in $X$ and $Z$ to always connect if they lie on the outer face. 
  Similar to the gadget for synchronizing variables, each of the pillars $V \in \{X,X', Y, Z\}$ has a degree-2 vertex $p_V$ on one side forming a wedge that encloses a vertex of $H_c$.
  As before, this construction allows us to ensure that the wedges of $X$ and $X'$ can never face each other.
  The wedge $p_Z$ of pillar $Z$ lies below all other wedges and, in contrast to the other pillars, the cluster $H_c$ does not span to the other side of $Z$.
  We refer to \Cref{fig:PM3SatOverview} for the exact structure of the clause gadget.
  
  \Cref{fig:PM3SatClauseGadgetInvalid} shows the one configuration of truth values for the three variables that does not allow an admissible embedding of the clause gadget.
  For both possible flips of $X'$, the $y$-monotonicity we require for the connections between cluster components prohibits a valid embedding of the gadget.
  Either the vertex of cluster $H_c$ that is enclosed by the wedge $p_Z$ in $Z$ cannot connect to any other vertex of $H_c$ (\Cref{fig:PM3SatClauseGadgetInvalidA}), or the cluster component contained in $X'$ cannot connect to the cluster component contained in $X$ (\Cref{fig:PM3SatClauseGadgetInvalidB}).
  In both cases, this configuration thus leads to an invalid embedding.
  For each of the other seven configurations, a valid embedding is possible; see \Cref{fig:PM3SatClauseGadgetValid}.

  Since the clause gadget enforces valid variable assignments and the literal gadget synchronizes the flip of corresponding literals, the resulting instance of \shortyclplan is a yes-instance if and only if the original instance of \pmsat was a yes-instance.
  Further, the resulting instance is single-source and, after adding an additional path on the outer face connecting the topmost and bottommost variables, the instance is also biconnected. \end{proof}

\threepartthm*

\begin{proof}

We describe a reduction from \textsc{3-Partition} to \shortyclplan with 5 levels.

\begin{figure}
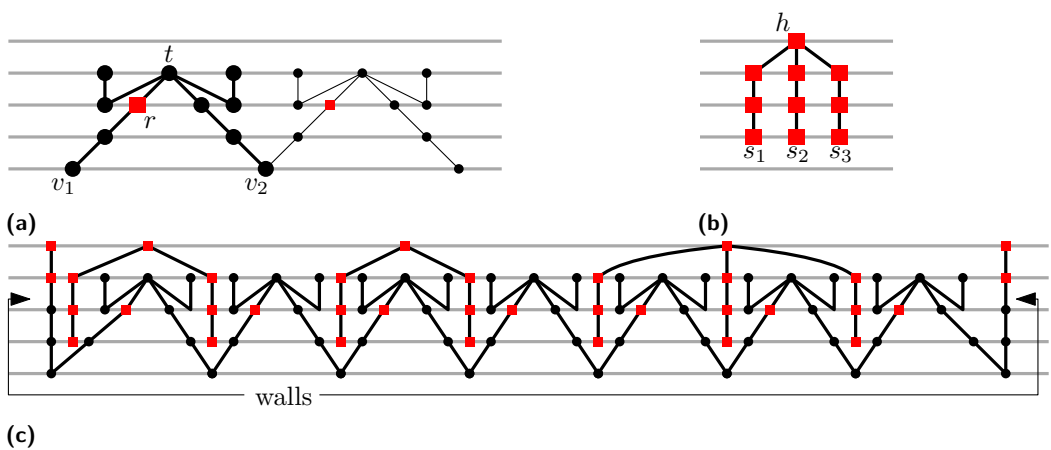

   \begin{subfigure}{.55\textwidth}
    \includegraphics[]{3part-receiver}
    \label{afig:receiver}
    \caption{}
   \end{subfigure}
   \begin{subfigure}{.35\textwidth}
    \includegraphics[]{3part-plug}
    \label{afig:plug}
    \caption{}
   \end{subfigure}
   \begin{subfigure}{\textwidth}
    \includegraphics[]{3part-bucket}
    \label{afig:bucket}
    \caption{}
   \end{subfigure}
    \caption{(a) A receiver (bold) with a marked vertex~$r$ (box), and a second receiver (non-bold) chained to the first one.
    (b) A 3-plug.
    (c) A bucket of size 7, filled with two 2-plugs and a 3-plug. Every marked bucket vertex can be connected to a pin with a $y$-monotone curve.}
    \label{afig:3part-gadgets}
\end{figure}

A \emph{receiver} is the structure described in \Cref{afig:3part-gadgets}(a). Its main component are its two endpoints $v_1, v_2$ on level 1, and a path connecting them that goes from level 1 to a vertex $t$ on level 4 and back. We mark the unique vertex on the path from $v_1$ to $t$, that lies on level 3 and call it $r$. Additionally, there are two paths of length 2, also called \emph{blockers} connected to $t$, that both have a vertex on level 3 and then go back to level 4. If the order of $v_1, v_2$ is fixed, \Cref{afig:3part-gadgets}(a) shows the unique level planar embedding (up to swapping the blockers) of a receiver. The two arms assure that the marked vertex $r$ can only be reached with a $y$-monotone curve from a vertex on level 2 or lower.
Two receivers can be chained by identifying $v_2$ of the first receiver with $v_1$ of the second receiver. A \emph{bucket} of size $B$ is obtained by chaining $B$ receivers in this manner, and then adding a path (also called \emph{wall}) to level 5 to each of the two unchained endpoints ($v_1$ of the first receiver, and $v_2$ from the last). A bucket contains $B$ marked vertices $r_1, r_2, \dots, r_B$, one for each of the chained receivers.
Like a receiver, the level planar embedding of a bucket is unique up to swapping of blockers and mirroring along the $y$-axis. Thus, every valley formed by two receivers or the first receiver with its wall contains precisely one marked vertex.
A \emph{plug} of size $k$ (also called a \emph{$k$-plug}) are $k$ paths from level 2 to level 4, and a vertex $h$ on level 5 that is connected to all path vertices on level 4, as shown in \Cref{afig:3part-gadgets}(b). We call the $k$ vertices on level 2 the \emph{pins} $s_1, s_2, \dots, s_k$ of the $k$-plug.
If a graph contains a plug and a bucket, then in every level-planar drawing of this graph the pin has to lie either completely inside the area bounded by the two walls, or completely outside of it (that is, either every vertex of the pin is drawn between the two wall-vertices of its level, or none of them).
Further, if a plug is drawn in a bucket, every pin of this plug is drawn in a valley of the chain, and the only marked vertex it can be connected to via a $y$-monotone curve is the (at most one) marked vertex of this valley. \Cref{afig:3part-gadgets}(c) shows a bucket of size 7 that contains two 2-plugs and one 3-plug, and in which every marked vertex of the bucket can be connected to a pin by a $y$-monotone curve.

Now let $(m, A, B)$ be an instance of \textsc{3-Partition} where $B$ is polynomial in $n$, and let $n = 3m$. 
We construct an instance $G$ of \shortyclplan. The graph consists of $m$ buckets and an $a_i$-plug for every $a_i \in A$. The clustering is given by tree $T$ with one inner vertex that has as children all plug vertices, all marked vertices of the receivers (in total $m\cdot B$) and the level 4 and 5 vertices of the bucket-walls. All of the other vertices in $G$ are children of the root in $T$. In other words, besides of the trivial cluster that contains all vertices, there is only one cluster $C$ that contains the plugs, the marked vertices, and the upper parts of the walls.
It is easy to verify that the size of this instance is bounded in $O(m\cdot B)$, which is polynomial in the size of the \textsc{3-Partition} problem. Thus, it can also be generated in polynomial time.
It is left to show that there exists a solution $A_1, A_2, \dots, A_m$ for $(m, A, B)$ if and only if $G$ is $y$-monotone cl-planar.

Assume $G$ is $y$-monotone cl-planar, and let $\mathcal{G}$ be a $y$-monotone cl-planar drawing of $G$.
Let $P_i$ be a bucket of $G$. Since the non-trivial cluster $C$ is connected, every marked vertex $t$ in $P_i$ is connected to the cluster with a $y$-monotone curve. It cannot be connected to a cluster-vertex of a wall since the arms obstruct this path. So the only cluster-vertices it can be connected to are pins. Since every bucket contains $B$ marked vertices there are at least $B$ pins present in $P_i$. This holds for all buckets, therefore there are precisely $B$ pins in $P_i$, and if a pin is in $P_i$, all pins of the same plug are as well.
This means there exists a partition $K_1, K_2, \dots, K_m$ of the plugs, such that the values corresponding to the plugs in every set sum up to $B$.
The sets $A_1, A_2, \dots, A_m$, where every $A_i$ contains the values corresponding to the numbers of the plugs in $K_i$, are a valid solution for $(m, A, B)$.

Now assume that there exists a valid solution $A_1, A_2, \dots, A_m$ for $(m, A, B)$.
We construct a $y$-monotone cl-planar drawing $\mathcal{G}$ for $G$, in which every bucket will be filled with plugs as in \Cref{afig:3part-gadgets}(c).
In the first step, we draw the buckets $P_1, \dots, P_m$ in this order from left to right next to each other, and for each $i \in \{1, \dots, m-1\}$ we connect the level-4 vertex of the left wall of $P_i$ and the level-5 vertex of the right wall of $P_{i+1}$ with a straight line (thus pairwise connecting these cluster components).
Now for every $i \in \{1, \dots m\}$ we draw the plugs corresponding to the set $A_i$ into the bucket $P_i$ from left to right, such that every valley with a marked vertex contains precisely one pin. The pins can now be connected with a marked vertex $t$ each, connecting every cluster component of a marked vertex to a plug cluster component. At the end, in every bucket we draw connections from the left wall to the first plug, the first to the second plug, the second to the third plug, and the third plug to the right wall (always with a straight line from a level-4 vertex to a level-5 vertex). In this manner, $C$ is connected with $y$-monotone curves in $\mathcal{G}$ and $\mathcal{G}$ is cl-level-planar, so $\mathcal{G}$ is a valid solution for $G$.

Clearly, this result holds even if the rotation system of the graph is fixed.
\end{proof}

\end{document}